\documentclass{revtex4}

\usepackage{latexsym}
\usepackage{amsfonts,amssymb,amsmath,amsthm}
\usepackage{natbib}

\begin{document}
\title{Divergences in the vacuum energy for frequency-dependent interactions}
\author{D. V. Vassilevich}
\affiliation{CMCC, Universidade Federal do ABC, Santo Andr\'e, SP, Brazil}
\affiliation{Department of Theoretical Physics, 
St.Petersburg University, Russia}
\begin{abstract}
We propose a method for determining ultra-violet divergences in the vacuum
energy for systems whose spectrum of perturbations is defined through a
non-linear spectrum problem, i.e, when the fluctuation operator itself depends
on the frequency. The method is applied to the plasma shell model, which 
describes some properties of the interaction of electromagnetic
field with fullerens. We formulate a scalar model, which simplifies the matrix
structure, but keeps the frequency dependence of the plasma shell, and
calculate the ultra-violet divergences in the case when the plasma sheet
is slightly curved. The divergent terms are expressed in terms of surface
integrals of corresponding invariants. 
\end{abstract}
\maketitle

\section{Introduction}
Recent advances in experimental techniques and in the nano-sciences have
caused an increase in the interest to the theory of the Casimir effect
\cite{Bordag:2001qi,Milton:2004ya}. The quantum field theory (QFT) methods
proved to be very efficient in the study of the Casimir effect.
However, real materials behave rather different from what we have got used
to in conventional QFT. The properties of these materials change drastically
at short distances and/or high frequencies. Numerically, the details
of this behavior may give a dominant contribution to the Casimir energy
\cite{Marachevsky:2001pc}. The main focus of the present paper will be
on the plasma sheet (or plasma shell) model introduced a few years ago
by Barton \cite{Barton1,Barton2,Barton3} to describe certain features of
interaction of giant carbon molecules (building blocks of fullerens) 
with the electromagnetic field.
The matching conditions for the electromagnetic field become in this
model frequency-dependent, and this changes drastically the spectrum
of the fluctuations.

The first problem one faces when applying the QFT methods to the Casimir
energy calculations is the evaluation of divergent terms and 
renormalization. Though calculations for selected simple configurations
may help to understand the behavior of the finite part of the Casimir
energy, the requirement of selfconsistency (renormalizability) 
demands precise form of the counterterms on considerably more general
backgrounds. This can be obtained relatively easy if the interaction with 
the background does no depend on the frequency, but becomes a tough
problem if it does. 

To understand the problem better let us start with a frequency-independent
case. Consider a system where the spectrum
of the fluctuations is defined by a linear equation 
\begin{equation}
D\varphi_\omega = \omega^2\varphi_\omega\,,\label{lsp}
\end{equation}
where $\omega$ is the frequency of the fluctuations, and $D$ is a Laplace
type operator depending on background fields or other external conditions,
as, e.g., boundaries. Let $\rho (\omega^2)$ be the spectral density of the
eigenvalue problem (\ref{lsp}). Then one can define the vacuum energy as
\begin{equation}
\mathcal{E}=\frac 12 \int_\mu^\infty d(\omega^2)\, (\omega^2)^{\frac 12} \,
\rho(\omega^2)\,.\label{ven}
\end{equation}
(We use the units such that $\hbar=c=1$). $\mu$ is the lower bound of the
spectrum. The eigenvalues $\omega^2$ are supposed to be real. The integral
(\ref{ven}) is typically divergent at the upper limit and has to be
regularized and renormalized. In many regularization schemes the divergences
are defined by the heat kernel coefficients $a_k(D)$ with $k\le n+1$, with
$n$ being the number of the spatial dimensions. The heat kernel coefficients
are the coefficients in a small $t$ asymptotic expansion of the heat
trace of the operator $D$,
\begin{equation}
K(D,t)={\rm Tr}\, \biglb( e^{-tD}\bigrb)
\simeq \sum_{k=0} t^{(k-n)/2} a_k(D)\,. \label{as}
\end{equation}
The coefficients $a_k$ are local invariants, which ensures locality of the
counterterms, and are known up to a rather high number $k$ for a very large
class of spectral problems \cite{GilkeyB,KirstenB,Vassilevich:2003xt}. 
Note, that the dependence of $D$ on the 
background fields may be practically arbitrary, as well as the shape of the
boundaries. The heat kernel expansion
defines therefore the ultra violet divergences at the one-loop order
for most of the relevant physical problems.

The situation becomes more complicated if the operator $D$ itself depends
on the frequency, i.e. if we are dealing with a {\textit{non-linear
spectral problem}}. If the dependence of $D$ on $\omega$ is rather mild,
as happens on stationary but non-static backgrounds
\cite{Fursaev:2000dv,Fursaev:2001yu,Fursaev:2002vi}, the divergences
can still be expressed through the asymptotics of the heat kernel
for $D$ where the dependence on $\omega$ is frozen. In time-space
noncommutative theories, which also lead to non-linear spectral
problems, the divergences of the vacuum energy are also
expressible through the polynomials of the background field, but
the structure of these polynomials becomes more complicated
\cite{Strelchenko:2007xh,Konoplya:2007xt}. However, it is unclear
whether this property holds in a more general case. For example,
one can calculate the divergences of the vacuum energy in a
dielectric with a frequency-dependent permittivity, but only 
for a dielectric body having the form of a ball, or of a cylinder,
or of a cube \cite{Falomir:2001uv,Bordag:2001fq,Barton:2001wd}.

Can one nevertheless find the asymptotic of spectral density
and the ultra-violet divergences for a fairly general class of 
geometries and background fields? The answer is positive, in principle
(i.e., modulo technical difficulties). The method we propose consists 
in the following. First one has to freeze the dependence of the operator
$D$ on frequency and construct the corresponding heat kernel, which
may have the form of an expansion in small variations $u_A$ of the
background fields, form of the boundary or boundary conditions, but
has to be exact for all values of $t$ and for all values of the 
parameters in $D$ which will later depend on the frequency. From this
heat kernel one calculates the spectral density for the frequency-independent
operator, which, in turn, defines the spectral density for the original
problem. The high-frequency behavior of the latter defines the
ultra-violet divergences of the vacuum energy. 

An important remark is in order. All expansions we shall actually consider
are {\textit{asymptotic}} expansions in local invariants constructed
from $u_A$ and their derivatives ordered according to the canonical
mass dimension. In the framework of such an expansion it does not
make much sense to compare actual numerical values of the invariants
or too look for non-local terms. The usefulness of such an expansion
is based on the observation that the counterterms in QFT are local.
On the other hand, application of the same expansion to calculations
of the finite part of the vacuum energy is risky, even though 
the expansions for the spectral densities may formally be integrated
as in (\ref{ven}). 

We test our method at the example of a slightly curved plasma sheet
(more precisely, at a scalar model which keeps essential features
of the TM model of the electromagnetic field interacting with the
plasma sheet). At the end, we find the part of the spectral density
which has the form of an integral of local quantities quadratic in
the extrinsic curvature, and which generates logarithmic divergences
of the vacuum energy. In the conventional QFT approach this divergent
part  defines local counterterms which must be added to the action.
In a more phenomenological approach it defines the behavior of the 
model near the cut off.

This paper is organized as follows. In the next section we describe
the method and study a simple example of a frequency-dependent mass
term. In section \ref{sec-ps} we discuss the plasma shell model, including 
the corresponding matching conditions, the heat kernel for a 
frequency-independent auxiliary problem, and the spectral densities with
their ultra-violet behavior. The last section contains a brief discussion
of the results and some future prospects.
\section{General strategy}\label{sec-ge}
Let us consider a non-linear spectral problem of the type
\begin{equation}
D(Q(\omega),u_A)\varphi = \omega^2\varphi,\label{nlsp}
\end{equation}
where $D$ is a second order partial differential operator
which depends on the frequency $\omega$ through a parameter
$Q(\omega)$. The parameter $Q$ may be present in the symbol 
of $D$ (as, e.g., a mass term), or enter the spectral problem 
through boundary or matching conditions (cf. sec.\ \ref{sec-ps}).
The parameters $u_A$ are supposed to be small. A good example
for $u_A$ is a background field. We suppose that the spectrum of $\omega^2$ 
is bounded from the below for reasonably small $u_A$. Next, we
assume that for all relevant values of $Q$ there exists the
heat trace
\begin{equation}
K(D(Q,u_A),t)={\rm Tr}\, \biglb( e^{-tD(Q,u_A)}\bigrb)\,.
\label{KDt}
\end{equation}
Let us expand $K(D(Q,u_A),t)$ in asymptotic series,
\begin{equation}
K(D(Q,u_A),t) = K_{(0)}(D(Q,u_A),t) + K_{(1)}(D(Q,u_A),t) 
+ K_{(2)}(D(Q,u_A),t)+
\dots \label{Kas}
\end{equation}
We shall consider exclusively the case when $u_A$ are fields having a
positive mass dimension. The expansion (\ref{Kas}) can then be arranged
according to the total mass dimension of the fields entering each term,
counting also the derivatives. I.e., $K_{(0)}$ does not depend on $u_A$,
$K_{(1)}$ is linear in $u_a$, $K_{(2)}$ contains the terms quadratic in $u_A$
and linear in the derivatives of $u_A$, etc. The structure of the expansion
(\ref{Kas}) is similar to the standard heat kernel expansion (\ref{as}),
but the numerical coefficients in front of local invariants constructed
from $u_A$ become functions of $Q$ and $t$. The existence of the
asymptotic expansion (\ref{Kas}), and the actual construction of such
am expansion, are the most non-trivial and demanding parts of our
procedure. 

The heat trace can be represented through a Laplace transform of the
spectral density $\rho (\lambda,Q;u_A)$,
\begin{equation}
K(D(Q,u_A),t)
=\int_\mu^\infty d\lambda\, e^{-\lambda t} \rho (\lambda, Q;u_A)\,,
\label{KlQ}
\end{equation}
where $\mu$ is the lower bound of the spectrum. The expansion (\ref{Kas})
defines through the inverse Laplace transformation an expansion of the spectral
density
\begin{equation}
\rho (\lambda,Q,u_A) = \rho_{(0)}(\lambda,Q,u_A) 
+ \rho_{(1)}(\lambda,Q,u_A) 
+ \rho_{(2)}(\lambda,Q,u_A)+
\dots \label{ras}
\end{equation}

Next, we have to construct a spectral density corresponding to the
initial eigenvalue problem (\ref{nlsp}). Consider the eigenvalue problem
\begin{equation}
D(Q,u_A)\varphi_{Q,\lambda}=\lambda \varphi_{Q,\lambda}
\label{aux}
\end{equation}
 with a fixed $Q$.
The density $\rho (Q,\lambda;u_A)$ corresponds to this eigenvalue problem,
i.e., $\rho (Q,\lambda;u_A)d\lambda$ is the number of solutions of (\ref{aux})
with the eigenvalues between $\lambda$ and $\lambda +d\lambda$. Of course,
the functions $\varphi_{Q(\omega),\omega^2}$ do solve the equation 
(\ref{nlsp}), but the density is {\textit{not}} simply 
$\rho (Q(\omega),\omega^2;u_A)$ (with respect to $d\omega^2$)
since the spacing between the eigenvalues changes. However, the number of the
eigenvalues up to certain $\omega^2=\lambda$ remains the same. 
Namely, if we define a counting function
\begin{equation}
N(\lambda,Q;u_a)=\int_\mu^\lambda \rho (\nu,Q;u_A)d\nu \label{Nl}
\end{equation}
then
\begin{equation}
N(\omega^2;u_A)=N(\omega^2,Q(\omega),u_A) \label{Nn}
\end{equation}
is the counting function for the non-linear spectral problem (\ref{nlsp}), and
\begin{equation}
\rho (\omega^2;u_A)=\frac{\mathrm{d}}{\mathrm{d}\omega^2}
N(\omega^2;u_A)
\label{rn}
\end{equation}
is the corresponding spectral density (with respect to $d\omega^2$).
This result also follows from a rather complete analysis of non-linear
spectral problems presented in Refs.\ 
\cite{Fursaev:2000dv,Fursaev:2001yu,Fursaev:2002vi}. 

The expansion (\ref{ras}) defines an asymptotic expansion of the spectral 
density of the non-linear spectral problem
\begin{equation}
\rho (\omega^2;u_A) = \rho_{(0)}(\omega^2,u_A) 
+ \rho_{(1)}(\omega^2;u_A) 
+ \rho_{(2)}(\omega^2,u_A)+
\dots \label{rnls}
\end{equation}
Consider now the limit $\omega^2\to\infty$. According to (\ref{ven})
the part of the spectral density which decays as $\omega^{-3}$ or slower
generates ultra-violet divergences of the vacuum energy. Corresponding
counterterms can be isolated in each order of the expansion in $u_A$,
and the renormalization procedure can be carried out. As in the normal
case, the counterterms are expected to be local invariants constructed
from $u_A$.

Note, that $\rho_{(i)}$ are only parts of the full spectral density. 
Therefore, $\rho_{(i)}$ must not be positive.

Let us give here a short summary of the proposed method.
If one likes to evaluate divergences of the vacuum energy for a system
those fluctuations are described by the equation (\ref{nlsp}) up to
a certain order in $u_A$, one has to make the following steps.
\begin{itemize}
\item Consider an auxiliary problem with the dependence of $Q$ on 
$\omega$ frozen and calculate the heat kernel for $D$ up to the desired
order in $u_A$ but exactly in $Q$ and $t$.
\item Calculate the corresponding spectral density and the counting 
function $N(\lambda, Q;u_A)$.
\item The function $N(\omega^2,Q(\omega);u_A)$ is then the counting
function for our initial problem, and the derivative of  
$N(\omega^2,Q(\omega);u_A)$ with respect to $\omega^2$ gives the
corresponding spectral density. The behavior of this spectral density
for large $\omega$ defines ultra-violet divergences of the vacuum
energy.
\end{itemize}
Note, that the first two steps above are most difficult from the technical
point of view, and these steps should be done for the auxiliary problem.
Therefore, what really counts for the complexity of the method is how
$D$ depends on $Q$ rather than how $Q$ depends on $\omega$.

If we are considering only local terms (in $u_A$) in the expansions
above, we would miss non-local divergences it they do exist. Non-local
divergences are, however, to exotic to consider this is a serious
drawback of the method.  
\subsection{Frequency-dependent mass term}\label{sec-ma}
As a toy model, let us consider a spectral problem in $n=3$ spatial
dimensions with the operator $D$ of the form
\begin{equation}
D=-\partial_i^2 +m^2+V(x),\label{Dtoy}
\end{equation}
where $V(x)$ is a smooth potential which falls of at the infinity
of $\mathbb{R}^3$, $m^2$ is a mass term, which will be made 
frequency-dependent later. We shall be interested in the terms 
depending linearly on the potential. The corresponding part of
the heat kernel can easily be found (see, e.g., \cite{Vassilevich:2003xt})
\begin{equation}
K_{(1)}(m^2,V,t)=-\frac {t^{-1/2}}{(4\pi)^{3/2}}
\int d^3x\, V(x) e^{-m^2t}\,.\label{K1toy}
\end{equation}
By using (\ref{KlQ}) and (\ref{Nl}) one immediately finds the corresponding 
spectral density and the counting function
\begin{eqnarray}
&&\rho_{(1)} (\lambda,m^2;V)= -\frac 1{8\pi^2} (\lambda -m^2)^{-1/2}
\theta (\lambda-m^2) \int d^3x\, V(x)\,,\label{rto}\\
&&N_{(1)} (\lambda,m^2;V)= -\frac 1{4\pi^2} (\lambda -m^2)^{1/2}
\theta (\lambda-m^2) \int d^3x\, V(x)\,,\label{nto}
\end{eqnarray}
where $\theta$ is the step function which simply tells that the lower
bound of the integration in (\ref{KlQ}) and (\ref{Nl}) is $\mu=m^2$.
Let us now consider examples of the dependence of $m^2$ on $\omega$.

\paragraph{$m^2$ does not depend on the frequency.} By expanding 
$(\lambda - m^2)^{-1/2}$ in (\ref{rto}) for large $\lambda=\omega^2$,
$(\lambda - m^2)^{-1/2}\simeq \omega^{-1}+\frac 12 m^2 \omega^{-3}+\dots$
we find two divergent contributions to the vacuum energy. 
The term $\int d^3x V$ diverges linearly, and the term with $\int d^3x\,
m^2V$ diverges logarithmically. This is a well-known result.

\paragraph{$m^2=\alpha \omega^2$}. Taking $\lambda=\omega^2$ we obtain
$(\lambda - m^2)=(1-\alpha)\omega^2$. Therefore, for $\alpha<1$
all spectral functions are as in the case above with $m^2=0$
up to a multiplier of $\sqrt{1-\alpha}$. For $\alpha>1$ the spectral
density is non-zero for negative $\omega^2$ meaning that the system becomes
unstable. The same conclusion may be reached by considering the eigenvalue
problem 
$\omega^2\varphi_\omega=(\alpha\omega^2-\partial_i^2 +V)\varphi_\omega$.

\paragraph{$m^2=\gamma \omega^4$}. For $\gamma >0$ the spectral density
is non-zero for $0\le \omega^2 \le 1/\gamma$, i.e. there are no high
frequencies in the spectrum at all. Consequently, there are no 
ultraviolet divergences proportional to the potential. For $\gamma<0$,
instabilities appear due to the imaginary frequencies with
$\omega^2<-1/|\gamma|$. 
\section{Curved plasma sheet}\label{sec-ps}
\subsection{Matching conditions}\label{sec-mc}
The plasma shell (or plasma sheet in the noncompact case) 
model was introduced by Barton in Refs.\ \cite{Barton1,Barton2}
where also the Casimir energies were explored.
The Casimir force between the plasma sheet and other surfaces
or molecules was calculated in \cite{BKM} by using the Lifshitz
formulae.
Further spectral properties of the plasma shell model
were studied for flat \cite{Bordag:2005qv}
and spherical \cite{Bordag:2008rc} geometries. In both cases,
the TM mode of the electromagnetic field causes more difficulties
than the TE mode. The TE problem is isomorphic to the scattering
on a repulsive frequency-independent delta-potential, which is a rather 
well understood problem. The study of the heat kernel expansion and thus
of the divergences of the vacuum energy for the delta-potential on
various surfaces started in Ref.\ \cite{Bordag:1998vs} with the case of
a sphere. Then the heat kernel for an arbitrary shape of the surface in an
arbitrarily curved space, with generic bulk potential and gauge fields
was obtained \cite{Gilkey:2001mj}. The spectral problem for TM mode 
is known to be not elliptic \cite{Bordag:2005qv,Bordag:2008rc}. Thus
even the definition of basic spectral functions become problematic.

Here we consider exclusively the TM mode, which,
for a spherical shell placed at $r=r_0$
may be reduced to a scalar field $\varphi$
with the following matching conditions \cite{Barton1}
\begin{eqnarray}
&&{\mathrm{discont}}\, (r\varphi_\omega) = -\frac {2q}{\omega^2} 
(\partial_r r\varphi_\omega),\nonumber\\
&&{\mathrm{discont}}\, (\partial_r r\varphi_\omega)=0\,,\label{dTM}
\end{eqnarray}
where ${\mathrm{discont}}\, (f)\equiv f(r=r_0+0)-f(r=r_0-0)$. 
$q$ is a constant depending on the properties of the material. Outside
the shell the modes $\varphi_\omega$ satisfy the Helmholtz equation
\begin{equation}
(\nabla^2+\omega^2)\varphi_\omega =0.\label{Helm}
\end{equation}
In the limit $r_0\to\infty$ the matching conditions (\ref{dTM})
reproduce that of the flat plasma sheet \cite{Barton2,Bordag:2005qv}.

To prepare for the discussion of the next subsection we need a bit more
elaborate notations. Denote by $M^+$ the exterior of the shell, and by $M^-$
-- the interior of the shell. Let $\Sigma$ be the (spherical) interface
of $M^+$ and $M^-$, and let $n^\pm$ be a unit normal to $\Sigma$ pointing
inward of $M^\pm$. Denote corresponding covariant
derivatives by $\nabla_n^\pm$.
When acting on scalars, $\nabla_n^+=\partial_r$, and $\nabla_n^-=-\partial_r$.
Then we can rewrite (\ref{dTM}) as 
\begin{equation}
-\left( \begin{array}{c} \nabla_n \varphi^+ \\
\nabla_n \varphi^- \end{array} \right)=
\left( \begin{array}{cc} Q+\sigma & -Q \\
-Q & Q-\sigma \end{array}\right)
\left( \begin{array}{c} \varphi^+ \\
 \varphi^- \end{array} \right)\,, 
 \label{ps-match}
\end{equation} 
where 
\begin{equation}
Q(\omega)=\omega^2/(2q),\qquad \sigma=1/r_0. 
\label{Qs}
\end{equation}
The dependence of $\varphi$ on $\omega$ is suppressed. 

We shall consider the case when the surface $\Sigma$ has the topology of 
a plane but is slightly curved. $M^+$ and $M^-$ are two parts of the bulk
manifold separated by $\Sigma$. The non-flatness of $\Sigma$ is measured 
by the extrinsic curvature tensor $L_{ab}$ which has two indices corresponding
to the directions tangential to $\Sigma$. Obviously, the extrinsic curvature 
of $\Sigma$ considered as a part of $M^+$ is minus the extrinsic curvature of
$\Sigma$ as a part of $M^-$, $L_{ab}^+=-L_{ab}^-$. We shall consider a scalar
field with the matching conditions (\ref{ps-match}) on $\Sigma$, where,
according to (\ref{Qs}), $Q$ is independent of the extrinsic curvature,
while $\sigma$ is supposed to be of the order of $|L_{ab}|$, but
independent of the frequency. Therefore, $Q$ will be treated 
non-perturbatively, while only the leading orders of $\sigma$ and $L_{ab}$ 
will be  retained in all spectral functions.

Note, that for an arbitrary curved boundary the electromagnetic field cannot
be separated in the TE and TM parts. Therefore, the model we are going
to analyze is an approximation to real physics. We hope that the most
essential features of the problem are nevertheless preserved by this
simplification.

\subsection{Heat kernel for the auxiliary model}\label{sec-hk}
The auxiliary model is obtained by simply freezing the frequency
dependence in the matching conditions (\ref{ps-match}), i.e. by assuming that
$Q$ is some constant. We are going to evaluate the localized
heat kernel
\begin{equation}
K(f,D,t)={\mathrm{Tr}}\, (f\exp (-tD))\, \label{auxhk}
\end{equation}
where $f$ is some smooth localizing (or smearing) function, and
$D$ is the standard Laplacian, $D=-\nabla^2$. 

The method we are going to use is (an extension of) 
the method of Gilkey \cite{Gilkey:1975iq,Branson:1990xp},
see \cite{GilkeyB,KirstenB,Vassilevich:2003xt} for an extensive overview
with further explanations. The method is
based on a somewhat paradoxical observation that solving a more general
problem may be much easier. We introduce an arbitrary curved metric
$g_{ij}$ on $M$ and generalize $D$ to be an arbitrary scalar Laplacian
\begin{equation}
D=-(g^{ij}\nabla_i \nabla_j + E), \label{genD}
\end{equation}
where $E(x)$ is a potential. We suppose that the metric is smooth across
$\Sigma$, so that $L_{ab}^+=-L_{ab}^-$. At some stage, when discussing
the doubling trick, see Eq.\ (\ref{L4.1}) below, 
we shall need more general matching
conditions
\begin{equation}
-\left( \begin{array}{c} \nabla_n \varphi^+ \\
\nabla_n \varphi^- \end{array} \right)=
\left( \begin{array}{cc} S^{++} & S^{+-} \\
S^{-+} & S^{--} \end{array}\right)
\left( \begin{array}{c} \varphi^+ \\
 \varphi^- \end{array} \right) 
 \label{genS}
\end{equation}
depending on four functions $S^{\pm\pm}$. The functions $S^{\pm\pm}$
have a positive mass dimensions and are, therefore, natural parameters 
of a perturbative expansion. However, $S^{\pm\pm}=0$ does not correspond
to a free propagation, but rather to two disjoint regions with 
Neumann boundary conditions on the surface which separates them.
On a side note, we remark that this explains the failure of the 
multiple reflection expansion in the case of these
matching conditions \cite{Bordag:2001ta}. Throughout this subsection
we shall suppose that the dimension of $M$ is arbitrary,
$n={\rm dim}\, M$. The last ($n$th) coordinate is supposed to be normal
to $\Sigma$, so that the notations introduced above for the normal
and tangential coordinates remain consistent. The spectral geometry of 
Laplace type operators with such matching conditions has been
analyzed in \cite{Gilkey:2002nv}. Basing on the analysis of that
paper we can write the following expression for the heat kernel
for the conditions (\ref{ps-match}) which is non-perturbative in
$Q$ and contains other invariants up to the canonical mass dimension
two:
\begin{eqnarray}
&&K(f,D,t)\simeq \frac 1{(4\pi t)^{(n-1)/2}}\int_{\Sigma} d^{n-1}x\,
\sqrt{h}\, \biglb[ 
\beta_0(z)f + t \biglb( \beta_1 (z) (L_{ab}^+ - L_{ab}^-)
(L_{ab}^+ - L_{ab}^-) f + \beta_2 (z) (L_{aa}^+-L_{aa}^-)^2 f
\nonumber\\
&&\qquad\qquad 
+ \beta_3(z) (f_{;n}^+-f_{;n}^-)(L_{aa}^+-L_{aa}^-)
+ \beta_4 (z) \frac 12 (f_{;nn}^++f_{;nn}^-)
+\beta_5 (z) \sigma (L_{aa}^+-L_{aa}^-) f 
+ \beta_6(z) \sigma (f_{;n}^+-f_{;n}^-) \nonumber\\
&&\qquad\qquad 
+\beta_7(z) Ef + \beta_8(z) Rf + \beta_9(z) R_{anan} f +
\beta_{10} (z) \sigma^2 f \bigrb) \bigrb] + 
K_{\rm vol} (f,D,t).\label{genhk}
\end{eqnarray}
Here $\beta_i(z)$ are unknown function
depending on $Q$ through the dimensionless combination
\begin{equation}
z=2Qt^{1/2}\,.\label{zQt}
\end{equation}
(The results of the calculations are summarized at the end of this
subsection).
$Q$ is supposed to be constant, but $\sigma$ is allowed to depend
on the coordinates on $\Sigma$. 
The volume part cannot depend on $Q$ or $\sigma$
and is thus given by the standard expression
\begin{equation}
K_{\rm vol} (f,D,t)=\frac 1{(4\pi t)^{n/2}}
\int_M d^nx \sqrt{g} f (1+ t(E+R/6)).\label{Kv}
\end{equation}
We stress again that only the invariants of canonical mass dimension
two or less are being kept. Only a few functions $\beta_i$ will be
needed to analyze the curved plasma sheet model. Others are included
for technical reasons.

Some explanations regarding the formula (\ref{genhk}) are in order.
$R_{ijkl}$ is the Riemann tensor for the metric $g_{ij}$, and $R$
is the corresponding scalar curvature. $h$ is the determinant of the
metric induced on $\Sigma$. All tensor indices are taken in a local
orthonormal frame, so that there is no distinction between upper and
lower indices. $n^\pm$ is a unit vector along a normal geodesics pointing 
inward $M^\pm$. The indices $a$, $b$ correspond to the vectors
tangential to $\Sigma$. Summation over the repeated indices is 
understood. Semicolon denotes the covariant derivative, e.g.,
$f_{;nn}^+\equiv \nabla_n^+\nabla_n^+f^+$. The smearing function
$f$ and the metric are smooth across $\Sigma$, consequently,
$f^+_{;n}=-f^-_{;n}$, $L_{ab}^+=-L_{ab}^-$, $f_{;nn}^+=f_{;nn}^-$.
Therefore, the notations used in (\ref{genhk}) are redundant,
but they facilitate the comparison to Ref.\ \cite{Gilkey:2002nv}
and allow to keep track of the symmetries more easily. Eq.\ (\ref{genhk})
contains all possible invariants up to the mass dimension two.
Note, that the heat kernel must be invariant under interchanging
the roles of $M^+$ and $M^-$. Since $\sigma$ is antisymmetric under
$M^+\leftrightarrow M^-$, this excludes the surface integral of 
$\sigma$ as well as all other (non-vanishing) invariants of the mass
dimension one. Since $\sigma$ is defined on the interface only, one
is not allowed to differentiate it with respect to the normal vector. 
Finally, $R_{abab}$ is not an independent invariant.

Here we like to reiterate the remark made already a couple of times above.
The expression (\ref{genhk}) is an asymptotic expansion in local invariant
ordered according to the canonical mass dimension. It has nothing to do with
actual (numerical) smallness of corresponding terms.

Let us start the calculations of the unknown functions $\beta_i(z)$.
By standard arguments (see, e.g., section 4.1 of Ref.\ 
\cite{Vassilevich:2003xt}) one can show that $\beta_i(z)$ do not depend
on $n$, i.e. the whole dependence on the dimensionality of the
manifold resides in the pre-factor $(4\pi t)^{-(n-1)/2}$. 
This fact is extremely important for the methods we use here.
As we shall see below, the so called conformal relations yield
some relations on $\beta_i$ with coefficients which are linear in $n$.
Since $\beta_i$ themselves do not depend on $n$, each of the 
conformal relations in fact produces two independent conditions on 
$\beta_i$'s.

The functions $\beta_0$, $\beta_4$, $\beta_6$ and $\beta_{10}$
can be computed by using Lemma 4.1 from \cite{Gilkey:2002nv}
which relates the heat trace on a manifold with boundary with
Robin boundary conditions to the heat trace on a ``doubled'' manifold
with transfer conditions on the interface. This Lemma states the following.
Let $M^\pm = M^0$ be a Riemannian manifold, and let $D^\pm=D^0$
be a scalar operator of Laplace type. Fix an angle $0<\theta<\pi/2$.
Let $S^{++}$ and $S^{+-}$ be arbitrary. Set
\begin{eqnarray}
&&S^{-+}:=S^{+-},\nonumber\\
&&S^{--}:=S^{++}+(\tan \theta - \cot \theta) S^{+-},\label{L4.1}\\
&&S_\phi :=S^{++}+\tan \theta S^{+-},\nonumber\\
&&S_\psi :=S^{++}-\cot \theta S^{+-}.\nonumber
\end{eqnarray}
Then
\begin{equation}
K(f,D,t)=K( \cos^2\theta f^+ + \sin^2\theta f^-,D^0_\phi,t)+
K(\sin^2\theta f^++\cos^2\theta f^-,D^0_\psi,t).\label{KKK}
\end{equation}
On the left hand side of the equation above we have the heat kernel
for the operator $D$ on $M$ subject to transfer conditions on the interface,
while on the right hand side there are two heat kernels on $M^\pm$
for the operator $D^0$ with Robin boundary conditions
\begin{equation}
(\nabla_n \phi + S_\phi)\phi \vert_{\partial M^0} = 0
\qquad \mbox{and} \qquad
(\nabla_n \psi + S_\psi)\psi \vert_{\partial M^0} = 0\,, \label{SpSp}
\end{equation}
respectively. $f^\pm$ is a restriction of $f$ to $M^\pm$.

To control the invariants appearing with $\beta_0$, $\beta_4$, $\beta_6$ 
and $\beta_{10}$ it is enough to consider the case of constant $\sigma$. 
Then,
\begin{equation}
\tan \theta = \frac {\sigma}Q + \sqrt{\frac {\sigma^2}{Q^2} +1} =
1+\frac {\sigma}Q +\frac {\sigma^2}{2Q^2}+O(\sigma^3)\label{tsQ}
\end{equation}
and
\begin{equation}
S_\phi = -\frac {\sigma^2}{2Q} + O(\sigma^3),\qquad
S_\psi = 2Q + \frac {\sigma^2}{2Q} + O(\sigma^3) \,.\label{SsQ}
\end{equation}

Another ingredient we need, is the heat kernel for the free Laplacian
$\Delta=-\partial_i^2$ subject to the Robin boundary
conditions at a plane boundary with a constant $S$ which was 
derived in \cite{Bordag:2001fj}
\begin{equation}
K(f,\Delta_S,t)=\frac 1{(4\pi t)^{(n-1)/2}} \int_{\partial M}
d^{n-1}x\, \sum_{p=0}^\infty \bigglb( 
\sum_{k=1}^\infty f^{(p)} (St^{1/2})^{p+k}\frac {2^{-p-1}}{
\Gamma \biglb( \frac {k+p}2 +1 \bigrb)}+
f^{(p)}t^{p/2} \frac {2^{-p-2}}{
\Gamma \biglb( \frac {k+p}2 +1 \bigrb)} \biggrb)\,, \label{Krob}
\end{equation}
where $f^{(p)}$ is the $p$th normal derivative of $f$. 
We do not write here a trivial volume part of the heat kernel. For each $p$ 
one can also write a closed expression in terms of the error function
\begin{equation}
{\mathrm{erf}\, (z)}=1-{\mathrm{erfc}\, (z)}=\frac 2{\sqrt{\pi}}
\int_0^ze^{-u^2}du\,.\label{erf}
\end{equation}
For example, for $p=0$ we have
\begin{equation}
K(f,\Delta_S,t)=\frac 1{2(4\pi t)^{(n-1)/2}} \int_{\partial M}
d^{n-1}x\ \biglb[ e^{S^2t} {\rm erf}\, (St^{1/2}) + e^{S^2t} - \frac 12
\bigrb] \,.\label{KRerf}
\end{equation}
In the formulae above, Eqs.\ (\ref{Krob}) and (\ref{KRerf}) we corrected
the $S$-independent terms of the corresponding formulae in 
\cite{Bordag:2001fj}. Such terms, being of the order of ``zero reflections''
were outside of the focus of the paper \cite{Bordag:2001fj}.

Let us stress, that as noted in \cite{Bordag:2001fj}, the expression
(\ref{KRerf}) is exact, i.e. it takes into account the possibility
of non-analytic contributions which may not be seen at power-series
expansions, see also \cite{Blinder,CJ}.

By collecting (\ref{L4.1}) - (\ref{KRerf}) together, we obtain,
\begin{eqnarray}
&&\beta_0(z)=\frac 12 [e^{z^2} ({\rm erf}\, (z) +1)]\nonumber\\
&&\qquad = \frac 12 \sum_{k=0}^\infty z^k \frac 1{
\Gamma \biglb( \frac {k}2 +1 \bigrb)}\,, \nonumber \\
&&\beta_4(z) = \frac 18 \sum_{k=0}^\infty z^k \frac 1{
\Gamma \biglb( \frac {k}2 +2 \bigrb)}\,,\label{b0bb}\\
&&\beta_6(z) = 2\beta_4(z) \,,\nonumber\\
&&\beta_{10}(z)=2\beta_0(z).\nonumber
\end{eqnarray}
Some terms in the expansions above can be checked against
the corresponding terms in \cite{Gilkey:2002nv}. These expansions are 
rapidly convergent for positive $z$. Therefore, it does not play a role 
whether one uses closed formulae in terms of the error function, or
an expanded form. Using expanded forms is a little bit easier.

To define $\beta_6$ and $\beta_7$ we use the product Lemma (see, e.g.,
sec. 4.1 of Ref.\ \cite{Vassilevich:2003xt})
which states that if $M=M_1\times M_2$, and also $D=D_1\otimes 1+
1\otimes D_2$, with $D_1$ and $D_2$ acting independently on functions
over $M_1$ and $M_2$, the resulting heat kernel factorizes 
into a product of the heat kernel for $D_1$ and the heat kernel for 
$D_2$. This means that if $f(x_1,x_2)=f_1(x_1)f_2(x_2)$, where $x_1$
and $x_2$ are coordinates on $M_1$ and $M_2$, respectively, 
\begin{equation}
K(f,D,t)=K(f_1,D_1,t)K(f_2,D_2,t)\,.\label{Kfac}
\end{equation}
Let us take $M_2$ being a curved manifold without a singular
surface, and $M_1$ being a flat manifold with a singular surface
and define some transfer conditions on it. Let $D_1$ and $D_2$ be scalar
Lapalcians, and let $E$ be independent of the coordinates
on $M_1$. Then, to the linear order of $E$ and the scalar curvature $R$
\begin{equation}
K(f_2,D_2,t)=\frac t{(4\pi t)^{-n_2/2}} \int dx_2 \sqrt{g_2} \biglb(
E+\frac 16 R\bigrb).\label{ER}
\end{equation}
Then, by comparing corresponding terms on both sides of
(\ref{Kfac}), we obtain
\begin{equation}
\beta_7(z)=\beta_0(z)\,,\qquad \beta_8(z)=\frac 16 \beta_0(z)\,.
\label{b7b8}
\end{equation} 

The remaining functions $\beta_1$, $\beta_2$, $\beta_3$, $\beta_5$ and
$\beta_9$ will be computed by using conformal properties of the operator
$D$. One can show that
\begin{equation}
\left. \frac {d}{d\epsilon}\right\vert_{\epsilon=0} {\rm Tr}\, \left[
\exp \left( -e^{-2\epsilon f}Dt \right)\right]=
-2t \frac {d}{dt} {\rm Tr}\, \left[ f \exp(-tD) \right].
\label{cr1}
\end{equation}
It is easy to figure out (or to look up in 
\cite{GilkeyB,KirstenB,Vassilevich:2003xt}) how the geometric
quantities entering the operator $D$, Eq.\ (\ref{genD}), transform  
the local scale transformation $D\to e^{-2\epsilon f}D$.
In particular, the transformation of the metric is the standard
Weyl rescaling, $g_{ij}\to e^{2\epsilon f}g_{ij}$. The transformation
of other quantities, as the connection and the potential, are less standard,
as they contain specific contributions which
ensure the homogeneous transformation law for $D$.
For our purposes, it is enough to consider the transformations vanishing 
on $\Sigma$, $f\vert_\Sigma=0$. Under this assumption, the variations
of relevant geometric invariants on the interface $\Sigma$ read 
\begin{eqnarray}
&&\left. \frac {d}{d\epsilon}\right\vert_{\epsilon=0}E
=\frac 14(n-2) (f_{;nn}^++f_{;nn}^-)-
\frac 18 (n-2) (L_{aa}^+-L_{aa}^-)(f_{;n}^+-f_{;n}^-) 
\,,\nonumber\\
&&\left. \frac {d}{d\epsilon}\right\vert_{\epsilon=0}R=
-(n-1)(f_{;nn}^++f_{;nn}^-) +\frac 12 (L_{aa}^+-L_{aa}^-)(f_{;n}^+-f_{;n}^-) 
\,,\nonumber \\
&&\left. \frac {d}{d\epsilon}\right\vert_{\epsilon=0}R_{anan}=
\frac 12 (n-1) (f_{;nn}^++f_{;nn}^-)
- \frac 14 (L_{aa}^+-L_{aa}^-)(f_{;n}^+-f_{;n}^-) \,,\label{ct1}\\
&&\left. \frac {d}{d\epsilon}\right\vert_{\epsilon=0}L_{ab}^\pm =
-\delta_{ab}f_{;n}^\pm\,.\nonumber
\end{eqnarray}
We must keep the matching conditions invariant under the local conformal
(scale) transformations. This is achieved by transforming $\sigma$
to compensate the variation of the connection in $\nabla_n$,
\begin{equation}
\left. \frac {d}{d\epsilon}\right\vert_{\epsilon=0}\sigma = 
\frac 14 (n-2) (f_{;n}^+-f_{;n}^-)\label{ct2}
\end{equation}
(cf \cite{Gilkey:2002nv}). $Q$ does not transform.

Let us collect the terms with $\frac 12 (f_{;nn}^++f_{;nn}^-)$ in
(\ref{cr1}). This gives the equation
\begin{equation}
\frac 12 (n-2) \beta_7(z) 
-2(n-1)\beta_8(z)
+(n-1)\beta_9(z)
=[(n-3) -z\partial_z]\beta_4(z)\,,\label{cv1}
\end{equation}
which, with the help of (\ref{b7b8}), may be simplified to
\begin{equation}
\frac 16 (n-4) \beta_0(z) 
+(n-1)\beta_9(z)
=[(n-3) -z\partial_z]\beta_4(z)\,.\label{cv2}
\end{equation}
The functions $\beta_i(z)$ do not depend on $n$. By taking $n=1$
we obtain the identity
\begin{equation}
\frac 12 \beta_0(z)= [2+z\partial_z]\beta_4(z)\,,\label{id1}
\end{equation}
which may serve as a consistency check. Taking $n=4$ yields the
value of $\beta_9$,
\begin{equation}
\beta_9(z)=\frac 13 (1-z\partial_z)\beta_4(z)=
\frac 1{24} \sum_{k=0} z^k \frac {1-k}{\Gamma \biglb( \frac {k}2 +2 \bigrb)}
\,.\label{b9}
\end{equation}

Next, lets us collect the terms with $\sigma(f_{;n}^+-f_{;n}^-)$
in (\ref{cr1}). This yields the equation
\begin{equation}
(1-n)\beta_5(z) + \frac 12 (n-2) \beta_{10}(z)
=[(n-3)-z\partial_z]\beta_6(z)\,.\label{cv3}
\end{equation}
Again, $n=1$ provides a consistency check. For $n=2$ we obtain
\begin{equation}
\beta_5(z)=[1+z\partial_z]\beta_6(z)=
\frac 14 \sum_{k=0} z^k \frac {1+k}{\Gamma \biglb( \frac {k}2 +2 \bigrb)}\,.
\label{b5}
\end{equation}

The only structure which has not been used yet in the conformal
relations is $(f_{;n}^+-f_{;n}^-)(L_{aa}^+-L_{aa}^-)$. It gives
the condition
\begin{equation}
\frac 18 (2-n)\beta_7(z)
+\frac 12 (n-1)\beta_8(z)
-\frac 14 \beta_9 (z)
+\frac 14 (n-2) \beta_5(z)
+2(1-n)\beta_2(z)-2\beta_1(z)=
[(n-3)-z\partial_z]\beta_3(z)\label{cv4}
\end{equation}
or
\begin{equation}
\frac 1{24} (4-n)\beta_0(z)
-\frac 14 \beta_9 (z)
+\frac 14 (n-2) \beta_5(z)
+2(1-n)\beta_2(z)-2\beta_1(z)=
[(n-3)-z\partial_z]\beta_3(z)\,.\label{cv5}
\end{equation}
Since the functions $\beta_i(z)$ do not depend on $n$, this equation
is equivalent to two independent conditions. However, we still have three
unknown functions $\beta_1$, $\beta_2$, and $\beta_3$.

To get one more relation between the unknown functions we shall use the
properties of the heat kernel under local scale transformations when the
smearing function is also transformed. It is a purely formal
computation \cite{Gilkey:1975iq,Branson:1990xp} to show that
\begin{equation}
\left. \frac {d}{d\epsilon}\right\vert_{\epsilon=0} 
a_{n-2} (e^{-2\epsilon f}F, e^{-2\epsilon f}D)=0,\label{cr2}
\end{equation}
where, similarly to (\ref{as}), $a_k(F,D)$ are the coefficients 
in an expansion of the smeared heat kernel $K(F,D,t)$.
One has to note that for each dimension $n$ of the manifold the
equation (\ref{cr2}) restricts just one heat kernel coefficient,
in contrast to (\ref{cr1}). Let $\beta^k_i$ be the coefficient
in front of $z^k$ in the Taylor expansion of $\beta_i(z)$,
\begin{equation}
\beta_i(z)=\sum_{k=0}^\infty z^k \beta^k_i\,.\label{bzb}
\end{equation}
Since $z\sim t^{1/2}$, cf (\ref{zQt}), each $\beta_i^k$ contributes
to $a_{k+3}$. I.e., Eq.\ (\ref{cr2}) gives restrictions on $\beta^k$
in the dimension $n=k+5$. Lets us collect the terms in (\ref{cr2})
proportional to $(f_{;n}^+-f_{;n}^-)(F_{;n}^+-F_{;n}^-)$. Since
now the smearing function is also transformed, such terms may
also appear from conformal transformations of the second
normal derivative of $F$:
\begin{equation}
\left. \frac {d}{d\epsilon}\right\vert_{\epsilon=0} 
(F_{;nn}^++F_{;nn}^-)= -\frac 52 (f_{;n}^+-f_{;n}^-)(F_{;n}^+-F_{;n}^-)
+\dots \label{ct3}
\end{equation}
(cf. Ref.\ \cite{Branson:1990xp}). Dots denote other terms, e.g. 
$\simeq f_{;nn}F$, which we do not use in this calculation. We have,
\begin{equation}
-\frac 54 \beta^k_4 - (n-1)\beta^k_3 +\frac 14 (n-2) \beta^k_6=0\,.
\label{629}
\end{equation}
Next, we remember that $n=k+5$ and use (\ref{b0bb}) to obtain
\begin{equation}
\beta_3^k=\frac 1{64}\, \frac {2k+1}{\Gamma \biglb( 
\frac k2 +3 \bigrb)}\,.\label{b3}
\end{equation}

Now we are ready to compute the remaining functions. Taking $n=1$ and $n=2$
in Eq.\ (\ref{cv5}) gives $\beta_1$ and $\beta_2$, respectively,
\begin{eqnarray}
&&
\beta_1(z)=\frac 1{192} \sum_{k=0}^\infty z^k \frac{(k+1)(2k+1)}{
\Gamma \biglb( \frac k2 +3 \bigrb)}\,,\nonumber\\
&&
\beta_2(z)=\frac 1{384} \sum_{k=0}^\infty z^k \frac{(k+1)(5k+13)}{
\Gamma \biglb( \frac k2 +3 \bigrb)}\,.
\label{b1b2}
\end{eqnarray}

For the convenience of the reader we summarize the results of calculations
of this subsection in Table \ref{betas} by providing the references to
corresponding equations. One can easily check that that all $\beta_i(z)$
are consistent with the results of \cite{Gilkey:2002nv} where several 
leading terms of the $z$-expansions for each $\beta$ were calculated.

\begin{table}
\caption{\label{betas} Summary of the calculations of $\beta_i(z)$.}
\begin{ruledtabular}
\begin{tabular}{|c|c|c|c|c|c|c|}
$\beta_i$: & $\beta_0$, $\beta_4$, $\beta_6$, $\beta_{10}$ &
$\beta_7$, $\beta_8$ & $\beta_9$ & $\beta_5$ & $\beta_3$ &
$\beta_1$, $\beta_2$ \\
\hline
Equation: & (\ref{b0bb}) &
 (\ref{b7b8}) & (\ref{b9}) & (\ref{b5}) & (\ref{b3}) &
(\ref{b1b2})\\
\end{tabular}
\end{ruledtabular}
\end{table}

The last step is to sum up the $z$-expansions for selected functions
$\beta_i(z)$ which are going to use below in sec.\ \ref{sec-cu}
\begin{eqnarray}
&&\beta_5(z) =\frac 14 \frac{\mathrm{d}}{\mathrm{d} z}
\left[ \frac 1z \biglb(e^{z^2}(\mathrm{erf}\, (z) +1) -1
\bigrb)\right] \,,\label{b5n}\\
&&\beta_1(z) = \frac 1{12} \,\beta_5(z) - \frac 7{192} \tilde \beta (z)\,,
\label{b1n}\\
&&\beta_2(z) = \frac 5{48} \,\beta_5(z) - \frac 7{384} \tilde \beta (z)\,,
\label{b2n}
\end{eqnarray}
where
\begin{equation}
\tilde \beta (z)=\frac{\mathrm{d}}{\mathrm{d} z}
\left[ \frac 1{z^3} \left( e^{z^2} ( \mathrm{erf}\, (z) +1) -1 
-\frac {2z}{\sqrt{\pi}} - z^2 \right)\right]\,.\label{bt}
\end{equation}
Note, that in principal the power series expansion, even though they
are convergent for all positive $z$, define the functions up to 
non-analytic terms which are exponentially small at $z\to 0$.
Nevertheless, since the expression for the Robin heat kernel
(\ref{KRerf}) is exact and does not include non-analytic
parts, the functions $\beta_i$ with $i=4,\dots,10$ also do not include
non-analytic parts, and their expressions in terms of the error function
are also exact (though not all of them are presented explicitly
here). However, to obtain $\beta_3$ the use of an expanded form 
was essential to write down the conformal relation (\ref{629}).
Therefore, $\beta_3(z)$ in principle can contain a non-analytic
part. Let us assume that the non-analytic part of $\beta_3(z)$
is zero, which is a very natural assumption. Then, there is no
non-analyticity in $\beta_1$ and $\beta_2$ which are defined by the other
functions through algebraic equations, and $\beta_1$ and $\beta_2$
are restored in a unique way as written above. 

\subsection{Spectral densities: plane case}
We have completed the calculation of the heat kernel, and it remains
to calculate the spectral densities. However, there is a subtlety in
application of our methods in the present case, which is related to
the treatment of unstable modes with negative $\omega^2$. To understand
the problem we start with the case of a plane plasma sheet.

One can easily see that flat ($\sigma=0$, $L_{ab}^\pm=0$) auxiliary
problem (with $Q$ positive and independent of $\omega$) admits surface modes
which decay as $e^{-2Q |x^n|}$ for positive $x^n$ and as
$-e^{-2Q |x^n|}$ for negative $x^n$. The Helmholtz equation (\ref{Helm})
gives the dispersion equation 
\begin{equation}
\omega^2 = p^2 -(2Q)^2\,, \label{adis}
\end{equation}
where $p$ is the momentum parallel to the interface $\Sigma$. 
This dispersion relation
corresponds to the excitations localized at the surface of the interface
and having a negative mass squared $-4Q^2$. Due to the presence of such 
tachionic excitations, the heat kernel has a part which grows for large
positive $t$. This part is defined by the growing part of $\beta_0(z)$,
cf. Eqs. (\ref{genhk}) and (\ref{b0bb}). In $n=3$ dimensions it reads
\begin{equation}
K_{\rm tachion}(f,D,t)= \frac 1{4\pi t} e^{4Q^2t} \int_\Sigma d^2x f\,.
\label{Kta}
\end{equation}  
For $f=1$ the heat kernel above is divergent, which is a consequence of
the translation invariance of the problem. Due to this invariance, the relevant
quantities are densities per unit area of the interface, as, e.g., the surface
Casimir energy density. Therefore, we omit the integral of the smearing 
function in Eq.\ (\ref{Kta}) thus obtaining a surface density of the heat
trace. The corresponding spectral density and the counting function
\begin{eqnarray}
&&\rho_{\rm tachion}(\lambda,Q)=\frac 1{4\pi} \theta (\lambda +4Q^2),
\label{rta}\\
&&N_{\rm tachion}(\lambda,Q)=\frac 1{4\pi} 
(\lambda+4Q^2)\theta (\lambda +4Q^2),
\label{Nta}
\end{eqnarray}
also have the meaning of surface densities. By using Eqs.\ (\ref{Nn}),
(\ref{rn}) and (\ref{Qs}), one obtains the part of the spectral density
for the nonlinear spectral problem, which is generated by the tachionic
part of the auxiliary problem:
\begin{equation}
\rho_{\rm tachion}(\omega^2)
=\frac 1{4\pi} \left(1+ \frac {2\omega^2}{q^2}\right)
\theta \left( \omega^2 + \frac {\omega^4}{q^2} \right)\,.
\label{rtt}
\end{equation}
This spectral density has two branches. One corresponds to real frequencies,
$\omega^2>0$, while the other - to imaginary frequencies, $\omega^2<-q^2$. 

The flat plasma sheet model can also be analyzed directly. There is a
mode, the so-called surface plasmon, which falls exponentially away from
$\Sigma$ with the eigenfrequencies \cite{Bordag:2005qv}
\begin{equation}
\omega^2_{\rm s.p.}=\frac q2 \left( \sqrt{q^2+4p^2}-q \right)\,.
\label{osp}
\end{equation}
The spectral density corresponding to the surface plasmon is simply
the density of the momenta $p$ parallel to $\Sigma$, i.e., 
$(2\pi)^{-2}d^2p$, which after integration over the directions of
$p$ becomes $(4\pi)^{-1} d|p|^2$. To obtain the spectral density in terms of 
$\omega^2$ one has to change variables in the previous expression which
yields
\begin{equation}
\rho_{\rm s.p.}=\frac 1{4\pi} 
\left( \frac {d\omega^2_{\rm s.p.}}{dp^2}\right)^{-1}\,.
\label{rsp}
\end{equation}
After some simple algebra the density (\ref{rsp}) can be shown to reproduce
the positive branch of the spectral density (\ref{rtt}). One can also 
demonstrate that the negative branch of (\ref{rtt}), $\omega^2<-q^2$,
correspond to the solutions which grow exponentially away from $\Sigma$
and thus do not belong to the physical spectrum. In our approach,
the negative branch must be excluded by hand. Obviously, in one perturbs
the plane plasma sheet by making is slightly curved, this cannot result
in the eigenvalues far away from the spectrum of non-perturbed problem.

\subsection{Spectral densities: curved case}\label{sec-cu}
Let us now proceed with the case of a curved plasma sheet. Let us restrict
ourselves to the case of flat ambient space with no external potentials,
so that $R_{ijkl}=E=0$. Next, we put the smearing function $f=1$. If we
suppose that the extrinsic curvature and $\sigma$ are well localized,
the surface integral in (\ref{genhk}) remains convergent (otherwise,
we can consider the surface density of the vacuum energy as described
in the subsection above). Let us remind, that $\sigma$ is supposed
to be of the same order as the extrinsic curvature. Therefore,
we arrive at the following second-order heat kernel
\begin{eqnarray}
&&K_{(2)}(D(Q),t)= \frac 1{(4\pi )}\int_{\Sigma} d^{2}x\,
\sqrt{h}\, \biglb[ 
 \beta_1 (z) (L_{ab}^+ - L_{ab}^-)
(L_{ab}^+ - L_{ab}^-)  + \beta_2 (z) (L_{aa}^+-L_{aa}^-)^2 
\nonumber\\
&&\qquad\qquad\qquad
+\beta_5 (z) \sigma (L_{aa}^+-L_{aa}^-) 
+\beta_{10} (z) \sigma^2  \bigrb]  .\label{K2}
\end{eqnarray}
(the 1st order heat kernel $K_{(1)}$ vanishes identically).
The functions which remain relevant for the calculations of this 
subsection are $\beta_1$, $\beta_2$, $\beta_5$ and $\beta_{10}$.
The numbering conventions do not look very natural, but, unfortunately,
any change of notations would inevitably bring in misprints. The 
present author kindly asks the reader for understanding.  

Let us start with the contribution of $\beta_{10}$, cf. Eq.\
(\ref{b0bb}). By using the identity
\begin{equation}
e^{4Q^2t} {\mathrm{erfc}}\, (2Qt^{1/2})=\frac 2\pi \int_0^\infty d\lambda\,
e^{-\lambda t} \frac {\mathrm{d}}{\mathrm{d}\lambda}
\arctan \left(- \frac {2Q}{\lambda^{1/2}}\right) \label{id}
\end{equation}
we arrive at the following expression for the counting function
\begin{eqnarray}
&&N_{10}(\lambda,Q)=\frac 1{4\pi} \int_\Sigma d^2x\, \sigma(x)^2 \cdot
\bar N_{10}(\lambda,Q)\,,\nonumber\\
&&\bar N_{10}(\lambda, Q)=2\theta (\lambda +4Q^2) +
\frac 2\pi \arctan \left(\frac {2Q}{\lambda^{1/2}}\right) \,.\label{N10}
\end{eqnarray}  
By using Eqs.\ (\ref{rn}) and (\ref{Qs}) one obtains an expression
for the spectral density of the non-linear spectral problem
\begin{eqnarray}
&&\rho_{10}(\omega^2)=\frac 1{4\pi} \int_\Sigma d^2x\, \sigma(x)^2 \cdot
\bar \rho_{10}(\omega^2)\,,\nonumber\\
&&\bar \rho_{10}(\omega^2)=2\delta (\omega^2)
+\frac q{\pi \omega (\omega^2+q^2)}\,.\label{r10}
\end{eqnarray}
We dropped another delta-function, $\delta(\omega^2+q^2)$,
since it corresponds to a non-physical part of the spectrum in
accordance with the previous subsection. The part of $\rho_{10}(\omega^2)$
which decays as $\omega^{-3}$ at large $\omega$, namely
\begin{equation}
\rho_{10}^{\rm div}(\omega^2)=\frac 1{4\pi^2}\, \frac q{\omega^3}
\int_\Sigma d^2x\, \sigma^2 \label{10d}
\end{equation}
generates a logarithmic divergence in the vacuum energy (\ref{ven}).

Let us now proceed with contributions of other invariants appearing in
(\ref{K2}). The inverse Laplace transformation of the functions
$\beta_5$, $\beta_1$, $\beta_2$ (see (\ref{b5n}) - (\ref{bt})) may be
performed by using e.g. the textbook \cite{LP}. The resulting expressions
for the spectral densities are becoming rather lengthy, and thus we
keep track of divergent contributions only. They read
\begin{equation}
\rho^{\rm div}(\omega^2)=
\frac 1{4\pi^2 }\, \frac q{\omega^3} \int_{\Sigma} d^{2}x\,
\sqrt{h}\, \biglb[ 
 \frac 1{12} (L_{ab}^+ - L_{ab}^-)
(L_{ab}^+ - L_{ab}^-)  + \frac 5{48} (L_{aa}^+-L_{aa}^-)^2 
+ \sigma (L_{aa}^+-L_{aa}^-) 
+ \sigma^2  \bigrb]  ,\label{rrd}
\end{equation}
where we have also included (\ref{10d}). This is the main result of this paper 
in what concerns the curved plasma sheet model. 

The formula (\ref{rrd}) defines all divergences of the vacuum energy in
in the plasma sheet model up to the quadratic order in the extrinsic
curvature. We like to remind that in sec. \ref{sec-mc} we argued that
$\sigma$ has to be considered as being of the same order as $L_{ab}$. 
We see, that there are no divergences in the linear order, and that
all divergences are logarithmic. The expansion in the extrinsic curvature
is an asymptotic one. Although it is enough to state the presence of
divergences of certain kind and the necessity of counterterms, it is not
enough to use the spectral densities
for calculations of the finite part of the vacuum
energy, as it is not clear how large or small the higher order corrections
could be. 

In principle, one cannot totally exclude that some other divergences will also
appear in the case of curved plasma sheet. Here we mean higher orders
of the extrinsic curvature, or with derivatives acting on $L_{ab}$
or $\sigma$. However, we find this very unlikely. Higher order invariants
are usually less divergent, and the quadratic terms already diverge only
logarithmically. 

It is instructive to compare our results to the ones obtained in 
\cite{Bordag:2008rc} for the spherical plasma shell. Since the volume of 
a two-sphere of the radius $r_0$ is proportional to $r_0^2$,
and the extrinsic curvature is inverse proportional to $r_0$, the expression
(\ref{rrd}) does not depend on the radius. The paper \cite{Bordag:2008rc}
found {\textit{linear}} divergences which do not depend on the the radius
and, therefore, correspond to the extrinsic curvature squared. (One should
take into account that our conventions for the heat kernel coefficients
differ from the ones used in \cite{Bordag:2008rc}.) This means that the 
divergences of plasma model are sensitive to the topology of the
interface surface $\Sigma$ in contrast to (most of) the frequency-independent
problems. This is an expected result, since to derive the part of the
spectral densities which generates divergences we needed the heat kernel
for all values of $t$, and not only the small $t$ asymptotics.

From the spectral densities we have calculated above one can derive the 
the heat kernel coefficients for the non-linear spectral problem.
This step is, however, unnecessary in our approach since the divergences
have already been calculated.

Let us remind that the electromagnetic field may be reduced to scalar fields
only for some specific geometries (including flat and spherical ones).
Therefore, the problem studied in this section is an idealization
of the full problem. However, the main complication from considering the
electromagnetic field directly is a rather nontrivial matrix structure
of the matching conditions, which makes the calculations
more involved, but not to the point where the technical difficulties
become unresolvable.
\section{Conclusions}
In this paper we suggested a method to evaluate the ultra-violet
behavior of spectral density and resulting divergences of the vacuum
energy for the problems with generic non-linear dependence on the spectral
parameter. The divergences are represented in a form similar to the heat
kernel coefficients, i.e. through integrals of local invariants associated
with the problem. The method, however, requires considerably more information
on the spectral properties of the operator involved that just the asymptotic
expansion of the heat kernel. One needs the whole heat kernel for all
values of $t$. Also, the parameter which depends on the frequency has to be
treated non-perturbatively. 

The method was tested at the plasma model of Barton, for which the divergences
of the vacuum energy for of a slightly curved plasma sheet were calculated
up to the second order in the extrinsic curvature. These divergences appeared
to be logarithmic, and they depend crucially on the topology of the
interface surface. This fact has certain implications for the plasma sheet 
model. Clearly it implies that a renormalizable model (in the sense of
\cite{Bordag:2008rc}) must contain some topology-dependent contributions
perhaps corresponding to global degrees of freedom.

We expect that the method can be considerably improved by using recent
advances in related areas. For example, to calculate the heat kernel one can
employ the worldline formalism \cite{Bastianelli:2008vh}. Recent advances
in calculation of the Casimir force between bodies of various shapes 
may also be useful (cf. \cite{Emig:2007qw,Bordag:2008gj,Wirzba:2007bv,
Dobrich:2008zz,Rey,Mar,Ahmedov:2008je} and references therein).
We also like to note some similarities to the method of Ref.\ 
\cite{BK} applied to one-dimensional systems with arbitrary
dispersion relations. Finally, we expect that with some modifications
the methods proposed above may be applied to such complicated
problems as boundary conditions depending on the tangential momentum
studied in Ref.\ \cite{Barvinsky:2006pg}.
\begin{acknowledgments}
I am grateful to Michael Bordag for fruitful discussions and warm hospitality
extended to me in Leipzig. This work was supported in part by CNPq.
\end{acknowledgments}


\end{document}